\documentclass[aip,reprint,groupedaddress]{revtex4-1}
\usepackage{graphicx}
\begin{document}

\title{Thermal insulator transition induced by interface scattering}

\author{Brian A. Slovick}
\author{Srini Krishnamurthy}
\affiliation{Applied Optics Laboratory, SRI International, Menlo Park, CA 94025, USA}
\begin{abstract}
We develop an effective medium model of thermal conductivity that accounts for both percolation and interface scattering. This model accurately explains the measured increase and decrease of thermal conductivity with loading in composites dominated by percolation and interface scattering, respectively. Our model further predicts that strong interface scattering leads to a sharp decrease in thermal conductivity, or an insulator transition, at high loadings when conduction through the matrix is restricted and heat is forced to diffuse through particles with large interface resistance. The accuracy of our model and its ability to predict transitions between insulating and conducting states suggest it can be a useful tool for designing materials with low or high thermal conductivity for a variety of applications.
\end{abstract}
\maketitle

The ability to obtain high or low thermal conductivity makes composite materials of great technological interest. Materials with low thermal conductivity are needed for insulating thermoelectrics (TE),\cite{Jones2006,Fleurial2009} while materials with high thermal conductivity are needed for thermal interface materials (TIM).\cite{Mahajan2004,Prasher2006} Since the efficiency of a TE device increases with the figure of merit $ZT$,\cite{Zhao2014} which is proportional to temperature, improving the insulation of a TE device leads to less lateral heat loss, larger operating temperatures, and thus higher efficiency.\cite{Fleurial2009} On the other hand, TIMs are needed to transfer heat from semiconductor devices such as transistors to a heat sink or heat spreader.\cite{Mahajan2004,Prasher2006,Cahill2014} Improving the thermal conductivity of TIMs reduces the thermal resistance of the heat transfer system and lowers the operating temperature of the underlying device, improving lifetime and reliability.

The current approaches for achieving high or low thermal conductivity have considerable limitations. For example, the porous silica aerogels\cite{Jones2006,Fleurial2009} used to insulate TE devices are mechanically brittle, leading to short lifetimes and high maintenance costs, while the thermal conductivity of TIMs consisting of ceramic particles in a polymer matrix is limited by thermal resistance associated with the mismatch in elastic properties at the interfaces.\cite{Mahajan2004,Prasher2006}

To improve the performance of insulators and TIMs, it is necessary to develop models that account for all of the physical phenomena that affect thermal conductivity. For composite materials, this includes two competing effects: percolation and interface scattering.\cite{Davis1975,Prasher2006,Prasher2006a,Tian2007} Percolation, which occurs at high loading when a continuous path is formed through the conductive phase, increases the effective thermal conductivity. Interface scattering, which arises from the difference in the phonon density of states between the particle and matrix,\cite{Swartz1989} reduces the effective thermal conductivity, and is particularly important for nanoparticles with large surface-to-volume ratio.

The most accurate models for calculating thermal conductivity of composite materials are based on molecular dynamics simulations or the Boltzmann transport equation.\cite{Yang2004,Cahill2014} While accurate, these models are computationally expensive, particularly for random mixtures. Alternatively, effective medium models derived from diffusion theory are simple and efficient, but inaccurate at high loading and the nanoscale because they do not account for both percolation and interface scattering. For example, the Maxwell-Garnett model assumes discontinuous discrete particles.\cite{Carson2005} As a result, it does not capture percolation and is accurate only for small volume loadings. For large loadings, the most widely used models are the symmetric \cite{Hakansson1990,Carson2005} and asymmetric \cite{Every1992,Prasher2006} Bruggeman models. The Bruggeman asymmetric model (BAM) includes interfacial thermal resistance, but does not include the effects of thermal percolation. Consequently, it underpredicts the thermal conductivity. Alternatively, the Bruggeman symmetric model (BSM) accounts for percolation, but does not include interfacial resistance, and thus overestimates the effective thermal conductivity.

In this Letter, we develop a generalized BSM (GBSM) that includes interface resistance and apply it to study thermal conduction in composite materials. Like the BSM, our model predicts a sharp increase in thermal conductivity at percolation when the interface scattering is weak. On the other hand, when the interface scattering is strong, our model also predicts a sharp decrease in thermal conductivity, or an insulator transition, at high loading. We also apply the model to explain the measured dependence of thermal conductivity on loading and particle size. The accuracy and simplicity of our model, and its ability to predict transitions between insulating and conducting states, suggest it can be a useful tool for designing materials with low or high thermal conductivity for a variety of applications.

To derive our model, we consider a spherical particle of thermal conductivity $\kappa_p$, embedded in a matrix of thermal conductivity $\kappa_m$, subjected to an external heat flux $q$. The solution of the heat equation for the temperature in the matrix ($T_m$) and the particle ($T_p$), relative to the temperature at the center of the particle, is \cite{Hasselman1987,Carson2005}
\begin{equation}
T_m=-\frac{q}{\kappa_m}r\cos{\theta}\left(1+\frac{A}{r^3}\right) \quad \text{and} \quad T_p=Br\cos{\theta},
\end{equation}
where $r$ is the distance from the center of the sphere, $q$ is along $z$ ($=r\cos{\theta}$), where $\theta$ is the angle between $r$ and $z$, and $A$ and $B$ are integration constants determined by the boundary conditions. Without interface resistance, the temperature and normal component of heat flux are continuous at the particle boundary, i.e., at $r=a$ \cite{Hasselman1987,Carson2005}
\begin{equation}
T_m=T_p \quad \text{and} \quad -\kappa_m \frac{\partial T_m}{\partial r}=-\kappa_p\frac{\partial T_p}{\partial r}.
\end{equation}
Applying Eq. (2) to Eq. (1), we obtain
\begin{equation}
A=a^3\frac{\kappa_m-\kappa_p}{2\kappa_m+\kappa_p} \quad \text{and} \quad B=-\frac{3 q}{2\kappa_m+\kappa_p}.
\end{equation}
The temperature in the matrix is then
\begin{equation}
T_m=-\frac{q}{\kappa_m}r\cos{\theta}\left[1+\frac{a^3}{r^3}\left(\frac{\kappa_m-\kappa_p}{2\kappa_m+\kappa_p}\right)\right],
\end{equation}
where the second term in Eq. (4) represents a local distortion in $T_m$ due to the particle. 

We now consider a composite containing a random distribution of spheres. In the BSM, the effective thermal conductivity $\kappa_{eff}$ of the composite is defined such that the volume average of the local distortion of $T_m$ is zero, which is equivalent to the condition \cite{Kirkpatrick1973,Davis1975,Carson2005}
\begin{equation}
\int{ \frac{\kappa_{eff}-\kappa}{2\kappa_{eff}+\kappa}p(\kappa})d\kappa=0,
\end{equation}
where $p(\kappa)d\kappa$ is the probability that the material in an arbitrary location has a conductivity $\kappa$. Since this probability is proportional to the volume fraction $f$, $\kappa_{eff}$ for a two-phase composite is thus given by \cite{Davis1975,Carson2005,Prasher2006}
\begin{equation}
 f\frac{\kappa_{eff}-\kappa_p}{2\kappa_{eff}+\kappa_p}+(1-f)\frac{\kappa_{eff}-\kappa_m}{2\kappa_{eff}+\kappa_m}=0.
\end{equation}
The solution of Eq. (6) is the BSM. It is invariant with respect to interchange of the particle and matrix, i.e., $\kappa_{eff}$ is unchanged when $\kappa_m\leftrightarrow\kappa_p$ and $f\rightarrow 1-f$. Therefore, the BSM provides a unique value of $\kappa_{eff}$ for a given set of materials, and does not distinguish between site (particle) and bond (matrix) percolation. We also note that the BSM can be generalized to multiphase composites by solving $\sum_i f_i(\kappa_{eff}-\kappa_i)/(2\kappa_{eff}+\kappa_i)$ with $\sum_i f_i=1$.

We now extend the BSM to include interfacial boundary resistance by imposing suitable boundary conditions, namely the heat flux is continuous at $r=a$, but the temperature is discontinuous by an amount proportional to the normal component of the heat flux at $r=a$,\cite{Hasselman1987}
\begin{equation}
T_m-T_p=R_b\kappa_{m,p}\frac{\partial T_{m,p}}{\partial r},
\end{equation}
where $R_b$ is the interfacial boundary resistance. Applying these boundary conditions to Eq. (1), we find that $A$ and $B$ have the same form as in Eq. (3), with $q$ replaced by $q/(1+R_b\kappa_p/a)$ and $\kappa_p$ replaced by \cite{Prasher2006,Hasselman1987}
\begin{equation}
\kappa'_p=\frac{\kappa_p}{1+R_b\kappa_p/a}.
\end{equation}
We find that $\kappa'_p$, which can be viewed as the effective thermal conductivity of a particle, decreases with the ratio of the boundary resistance to the thermal resistance of the particle ($a/\kappa_p$), and thus decreases with surface-to-volume ratio $1/a$. The closed-form solution of our model, obtained by solving Eq. (6) with $\kappa_p\rightarrow \kappa_p'$, is given by
\begin{eqnarray}
&&\kappa_{eff}=\frac{1}{4}[ K +(K^2 + 8\kappa'_p \kappa_m)^{1/2} ], \quad \text{where} \nonumber \\
&&K=\kappa_m(2-3f)+\kappa'_p(3f-1).
\end{eqnarray}

The evaluation of $\kappa_{eff}$ depends on the evaluation of $R_b$, which can be calculated from the net diffuse phonon flux transmitted across a semi-infinite interface due to a local temperature gradient,\cite{Swartz1989,Lombard2015}
\begin{eqnarray}
\frac{1}{R_b}=&&\frac{1}{2}\sum_j v_{m,j} \int_0^{\pi/2} \int_0^{\omega^{\text{max}}_{m,j}} d\omega D_{m,j}(\omega) \frac{\partial f(\omega,T)}{\partial T}\hbar \omega  \nonumber\\
&& \times  \tau_{m\rightarrow p}(\theta,j,\omega) \cos(\theta) \sin(\theta) d\theta,
\end{eqnarray}
where $j$ denotes the phonon polarization, $\omega$ and $\omega^{\text{max}}_{m,j}$, respectively, are the phonon frequency and maximum phonon frequency in the matrix, $\tau_{m\rightarrow p}$ is the transmission coefficient from the matrix to the particle, $v_{m,j}$ is the velocity of phonons in the matrix, $\hbar$ is Planck's constant, $D_{m,j}(\omega)$ is the density of states of the matrix, $f(\omega,T)$ is the Bose-Einstein distribution, where $T$ is the temperature, and $\theta$ is the angle of incidence. To obtain a closed-form expression for $R_b$, we make several approximations. First, we assume isotropic materials with linear dispersion (i.e., Debye solids), and thus $D_{m,j}(\omega)=\omega^2/(2\pi^2 v_{m,j}^3)$ and $\omega^{\text{max}}_{m,j}=v_{m,j} (6\pi^2 n_m)^{1/3}$, where $n_m$ is the atomic density of the matrix. Second, for matrix materials with low Debye frequencies such as polymers, $f(\omega,T)\approx k_BT/(\hbar \omega)$ at high temperatures, where $k_B$ is Boltzmann's constant. Lastly, since the dominant phonon wavelengths at high temperatures are small compared to the interface roughness, there will be scattering at the interfaces. Thus, $\tau_{m \rightarrow p}$ can be accurately calculated using the diffuse mismatch model \cite{Swartz1989,Lombard2015}
\begin{equation}
\tau_{m \rightarrow p}=\frac{v_{p,L}^{-2}+2v_{p,T}^{-2}}{v_{m,L}^{-2}+2v_{m,T}^{-2}+v_{p,L}^{-2}+2v_{p,T}^{-2}},
\end{equation}
where the subscripts $L$ and $T$, respectively, denote longitudinal and transverse phonon polarizations. With these approximations, Eq. (10) reduces to \cite{Lombard2015}
\begin{equation}
\frac{1}{R_b}=\frac{1}{4}  k_B n_m \tau_{m\rightarrow p}  (v_{m,L}+2v_{m,T}).
\end{equation}
Our model is completely contained in Eqs. (8), (9), (11), and (12), with all symbols defined in Table 1.

\begin{table}
\caption{Definition of symbols used in the model.}
\begin{ruledtabular}
\begin{tabular}{l c c c c}
Symbol & Property\\
\hline
$\kappa_m$ ($\kappa_p$)  & Thermal conductivity of the matrix (particle) \\
$f$  & Volume loading fraction of particles \\
$a$  & Radius of particles \\
$n_m$  & Atomic density of the matrix \\
$v_{m,L}$ ($v_{p,L}$)  & Velocity of longitudinal acoustic phonons \\
  & in the matrix (particle) \\
$v_{m,T}$ ($v_{p,T}$)  & Velocity of transverse acoustic phonons \\
  & in the matrix (particle) \\
\end{tabular}
\end{ruledtabular}
\end{table}

The effective thermal conductivity given by our model depends on the competing effects of percolation and interface scattering. Percolation increases the effective thermal conductivity of the composite, while interface scattering reduces the effective thermal conductivity of the particle ($\kappa_p'$). Since $\kappa'_p$ depends on the surface-to-volume ratio $1/a$, the effective thermal conductivity is critically dependent on radius. This is illustrated by Fig. 1, which shows $\kappa_{eff}$ calculated with our GBSM (solid lines) as a function of loading for three particle sizes with $\kappa_p/\kappa_m=100$. Also shown is $\kappa_{eff}$ calculated with the BAM (dashed lines). For a particular radius known as the Kapitza radius ($a_K$), the reduction in the effective thermal conductivity due to interface scattering exactly cancels the increase arising from percolation, and $\kappa_{eff}$ is independent of loading. The Kapitza radius, obtained by setting $\kappa_p'=\kappa_m$ in Eq. (8), is given by
\begin{equation}
a_K=\frac{R_b\kappa_p}{\kappa_p/\kappa_m-1}.
\end{equation}
For large particles with negligible interface scattering (i.e., $a/a_K\rightarrow \infty$), $\kappa_p'=\kappa_p$ and $\kappa_{eff}$ increases with loading. Alternatively, for small particles with strong interface scattering (i.e., $a<<a_K$), $\kappa_p'<\kappa_m$ and $\kappa_{eff}$ decreases with loading. While the BAM also predicts this general behavior, our GBSM predicts an abrupt change near percolation. In composites with weak interface scattering, the abrupt increase in $\kappa_{eff}$ at 30\% loading is well documented.\cite{Prasher2006,Tian2007} It occurs when a continuous path of conduction is formed between conductive particles. At the other extreme, when the interface scattering is strong, the particles have near-zero effective thermal conductivity, and conduction is largely through the matrix. However, owing to the symmetry of our GBSM, we find that $\kappa_{eff}$ decreases abruptly near 70\% particle loading, or when the loading of the more conductive matrix phase falls below the percolation value of 30\%. In this case, continuous paths of conduction through the matrix are restricted, and the heat is forced through contiguous particles with large interface resistance, resulting in ultralow effective thermal conductivity. This can be seen explicitly by taking the limit of our model as $a/a_K\rightarrow 0$, in which case $\kappa'_p=0$ and $\kappa_{eff}=0$ for $f\ge2/3$.

\begin{figure}
\includegraphics[width=51mm]{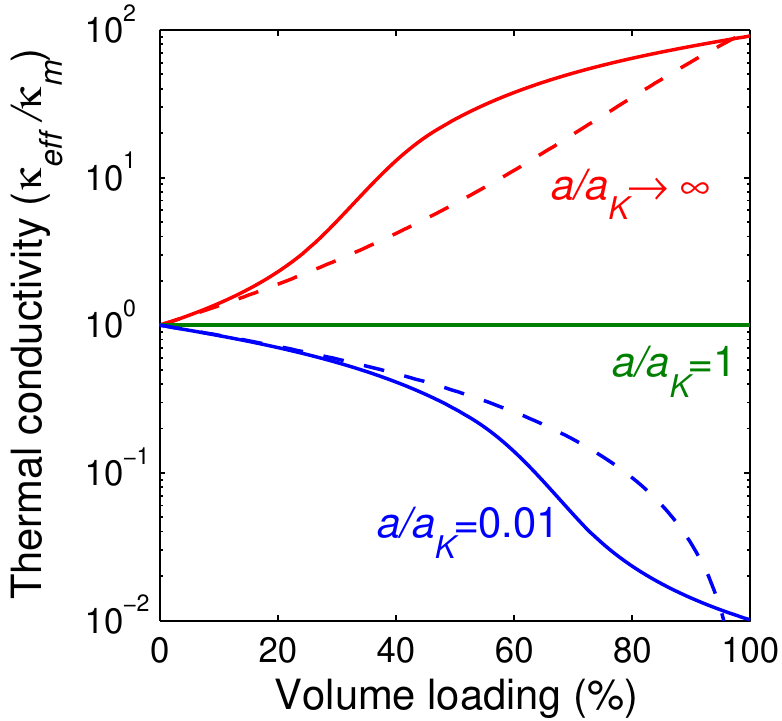}
\caption{\label{fig:epsart} Effective thermal conductivity relative to $\kappa_m$ versus volume loading of particles with $\kappa_p/\kappa_m=100$ and various sizes normalized to the Kapitza radius ($a_K$). The solid and dashed lines, respectively, show our GBSM and the BAM.}
\end{figure}

\begin{figure}
\includegraphics[width=51mm]{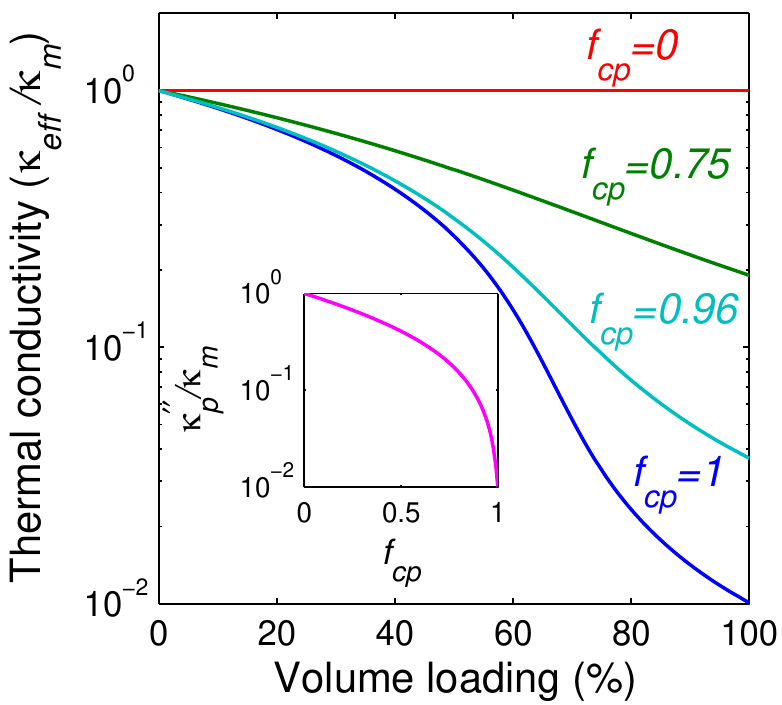}
\caption{\label{fig:epsart} Effective thermal conductivity as a function of volume loading of matrix-coated particles with $\kappa_p/\kappa_m=100$ and $a/a_K=0.01$, for various values of $f_{cp}$, the ratio of the volume of the particle to the volume of the coated particle. The inset shows the effective thermal conductivity of the coated particle as a function of $f_{cp}$.}
\end{figure}

Our model assumes that the interface is preserved for all loadings, whereas in practice, particle agglomeration at high loading may reduce the interfacial area and diminish the insulator transition. In principle, this can be avoided by coating the particles with a thin layer of the matrix material. To investigate the impact of a coating layer on the insulator transition, we used the Maxwell-Garnett equation to calculate the effective thermal conductivity of coated particles with interface scattering,\cite{Hasselman1987}
\begin{equation}
\kappa_p''=\kappa_m\frac{\kappa_p'+2\kappa_m+2f_{cp}(\kappa_p'-\kappa_m)}{\kappa_p'+2\kappa_m-f_{cp}(\kappa_p'-\kappa_m)},
\end{equation}
where $f_{cp}$ is the ratio of the volume of the particle to the volume of the coated particle. Figure 2 shows the thermal conductivity of the composite with $a/a_K=0.01$ from Fig. 1, as a function of loading for several values of $f_{cp}$. The inset shows $\kappa''_p$ as a function of $f_{cp}$. As $f_{cp}$ decreases from 1 (our original model), the effective thermal conductivity increases, as a larger fraction of the coated particle volume is occupied by the more conductive matrix coating. Therefore, as long as the coating layer is thin compared to the particle size (i.e., $f_{cp}\approx 1)$, the effective conductivity of the coated particle is dominated by the interface resistance and $\kappa_{eff}$ remains low.

We now apply our GBSM to understand the measured thermal conductivity of composites containing aluminum nitride (AlN) particles in epoxy \cite{Xu2000,Hsieh2006} and diamond particles in zinc sulfide (ZnS).\cite{Every1992} The material properties used in the calculations are shown in Table 2. The atomic density of epoxy was calculated from the specific heat ($c_m=1,060$ J/kg/K)\cite{Ganguli2008} and density ($\rho_m=1,150$ kg/m$^3$)\cite{Lindrose1978} using the Dulong-Petit relation $n_m=\frac{1}{3} \rho_m c_m/k_B$, and for $\kappa_p$ we used typical polycrystalline values. Figure 3a shows our GBSM and the measured thermal conductivity (circles) as a function of loading for 7 $\mu$m diameter AlN particles in epoxy ($R_b=0.064$ mm$^2$K/W).\cite{Xu2000} Also shown is the $\kappa_{eff}$ predicted by the BAM with $R_b=0$ and the BSM. In this case, the particle size is larger than the Kapitza diameter (26 nm) and $\kappa_{eff}$ increases with loading. Because the BAM does not account for percolation, it underpredicts the thermal conductivity for large loadings, even with $R_b=0$. On the other hand, the BSM (GBSM with $R_b=0$) overpredicts $\kappa_{eff}$ because it does not account for interface scattering. Our GBSM, which includes both effects, shows considerable agreement with the data, particularly with regard to the onset of percolation at 30\% and the linear increase of $\kappa_{eff}$ beyond percolation. Figure 3b shows the modeled and measured $\kappa_{eff}$ as a function of AlN diameter for a loading of 50\%.\cite{Hsieh2006} Our GBSM agrees well with the data, while the BAM with $R_b=0$ and BSM, respectively, underpredict and overpredict $\kappa_{eff}$. A plausible explanation for the relatively small overestimate of $\kappa_{eff}$ by our GBSM is that our model does not include interfacial resistance associated with imperfect chemical bonding between the particle and matrix. Also, we find that a GBSM based on the acoustic mismatch model \cite{Swartz1989} cannot explain the overestimate (not shown). Figure 3c shows the measured and modeled thermal conductivity for 100 nm diamond particles in ZnS ($R_b=0.013$ mm$^2$K/W).\cite{Every1992} In this case, the particle size is smaller than the Kapitza diameter (612 nm) and $\kappa_{eff}$ decreases with loading. While our GBSM captures the correct general trend, it overpredicts $\kappa_{eff}$. This is also likely due to poor bonding between the particle and matrix. To determine whether a large $R_b$ due to poor bonding can explain the data, we calculated the thermal conductivity using our GBSM and the BAM with $R_b\rightarrow \infty$, and find that both models closely follow the data up to 30\% loading.

\begin{table}
\caption{Material properties used in the calculations.}
\begin{ruledtabular}
\begin{tabular}{l c c c c}
Material & Thermal & Longitudinal & Transverse & Atomic\\
 &conductivity& speed & speed  & density\\
 &(W/m/K)& (m/s) & (m/s) & (1/m$^3$) \\
\hline
Epoxy \cite{Lindrose1978} & 0.2 & 2,377 & 1,128 & 2.9$\times 10^{28}$ \\
AlN \cite{McNeil1993} & 60 & 11,230 & 6,187 &  -\\
ZnS \cite{Every1992,Cline1967} & 17.4 & 5,510 & 2,640 &  5.0$\times 10^{28}$\\
Diamond \cite{McSkimin1957} & 600 & 17,500 & 12,800 &  -\\
\end{tabular}
\end{ruledtabular}
\end{table}

In summary, we developed an effective medium model of thermal conductivity for composite materials that accounts for both percolation and interface scattering. We applied our model to explain the measured dependence of thermal conductivity on loading and particle size. Depending on the ratio of the interface resistance to the thermal resistance of the particle, the effective thermal conductivity of the composite can be increased or decreased relative to the matrix. In particular, we showed that strong interface scattering can lead to an insulator transition at high loadings (67\%) when conduction paths through the matrix are eliminated. The accuracy and simplicity of our model, and its ability to predict transitions between insulating and conducting states, suggest it can be a useful tool for designing materials with low or high thermal conductivity for a variety of applications.

\begin{figure}
\includegraphics[width=54mm]{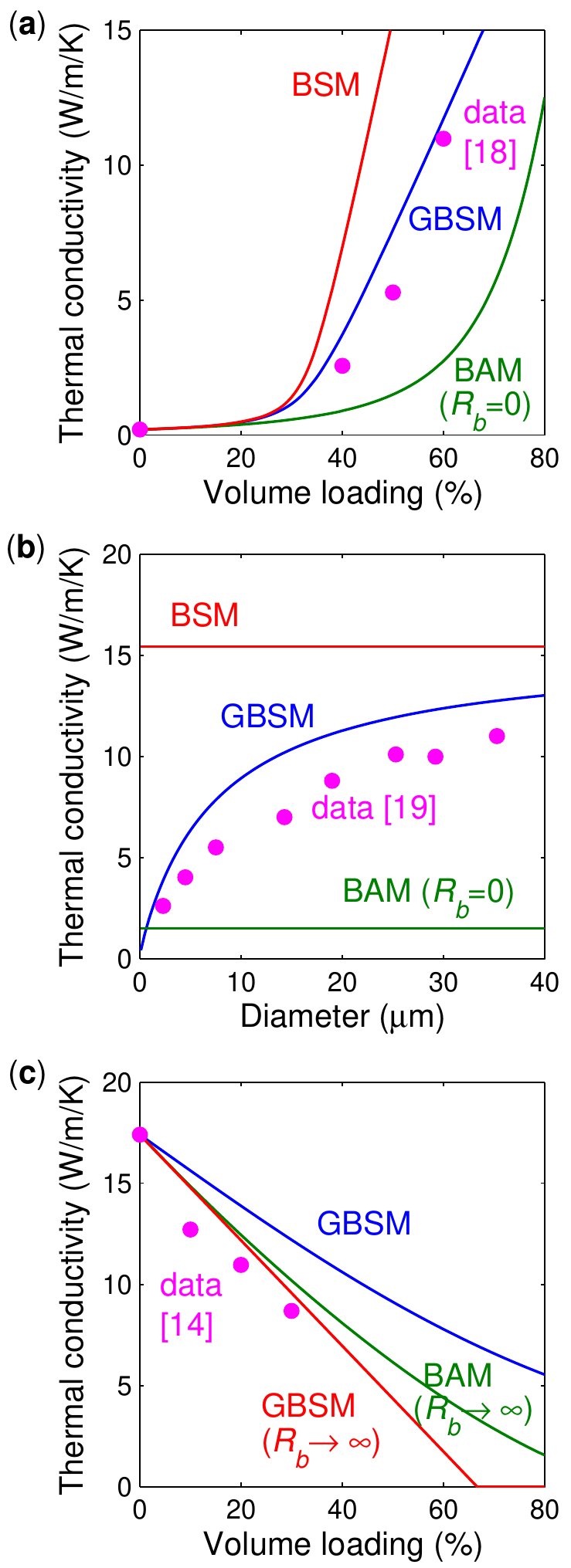}
\caption{\label{fig:epsart} (\textbf{a}) Thermal conductivity versus volume loading for 7 $\mu$m diameter AlN particles in epoxy. (\textbf{b}) Thermal conductivity versus diameter of AlN in epoxy for a loading of 50 \%. (\textbf{c}) Thermal conductivity versus volume loading for 100 nm diamond particles in ZnS.}
\end{figure}

\bibliography{bib}

\end{document}